\newcommand{\Ni}{\mbox{$^{56}$Ni}}
\newcommand{\Co}{\mbox{$^{56}$Co}}

\newcommand{\msol}{\mbox{M$_{\odot}$}\ }

\newcommand{\kms}{\mbox{$\rm{km}\,s^{-1}$}\ }

\newcommand{\degree}{\mbox{$^\circ$}}

\newcommand{\kmsMpc}{\,$\mathrm{km\,s}^{-1}\,\mathrm{Mpc}^{-1}$}
\newcommand{\ergs}{\,erg\,s$^{\mathrm{-1}}$}

\documentclass[twocolumn]{aastex631}

\shorttitle{SN 2022jli}
\shortauthors{Moore et al.}
\graphicspath{{./}{figures/}}
\begin{document}

\title{SN 2022jli: a type Ic supernova with periodic modulation of its light curve and an unusually long rise}

\correspondingauthor{T. Moore}
\email{tmoore11@qub.ac.uk}

\author[0000-0001-8385-3727]{T. Moore}
\affiliation{Astrophysics Research Centre, School of Mathematics and Physics, Queen's University Belfast, BT7 1NN, UK}

\author[0000-0002-8229-1731]{S. J. Smartt}
\affil{Department of Physics, University of Oxford, Denys Wilkinson Building, Keble Road, Oxford OX1 3RH, UK}
\affil{Astrophysics Research Centre, School of Mathematics and Physics, Queen's University Belfast, BT7 1NN, UK}

\author[0000-0002-2555-3192]{M. Nicholl}
\affil{Astrophysics Research Centre, School of Mathematics and Physics, Queen's University Belfast, BT7 1NN, UK}

\author[0000-0003-4524-6883]{S. Srivastav}
\affil{Astrophysics Research Centre, School of Mathematics and Physics, Queen's University Belfast, BT7 1NN, UK}

\author[0000-0002-0504-4323]{H.~F. Stevance} 
\affil{Department of Physics, University of Oxford, Denys Wilkinson Building, Keble Road, Oxford OX1 3RH, UK}
\affil{Department of Physics, The University of Auckland, Private Bag 92019, Auckland, New Zealand}

\author[0000-0002-9155-8039]{D.~B. Jess}
\affiliation{Astrophysics Research Centre, School of Mathematics and Physics, Queen's University Belfast, BT7 1NN, UK}
\affiliation{Department of Physics and Astronomy, California State University Northridge, Northridge, CA 91330, USA}

\author[0000-0001-5170-9747]{S.~D.~T. Grant}
\affiliation{Astrophysics Research Centre, School of Mathematics and Physics, Queen's University Belfast, BT7 1NN, UK}

\author[0000-0003-1916-0664]{M. D. Fulton}
\affil{Astrophysics Research Centre, School of Mathematics and Physics, Queen's University Belfast, BT7 1NN, UK}

\author[0000-0003-2705-4941]{L. Rhodes}
\affil{Department of Physics, University of Oxford, Denys Wilkinson Building, Keble Road, Oxford OX1 3RH, UK}

\author[0000-0002-9774-1192]{S. A. Sim}
\affil{Astrophysics Research Centre, School of Mathematics and Physics, Queen's University Belfast, BT7 1NN, UK}

\author[0000-0002-8032-8174]{R.~Hirai}
\affiliation{OzGrav: The Australian Research Council Centre of Excellence for Gravitational Wave Discovery, Clayton, VIC 3800, Australia}
\affil{School of Physics and Astronomy, Monash University, VIC 3800, Australia}

\author[0000-0002-8032-8174]{P.~Podsiadlowski}
\affiliation{University of Oxford, St Edmund Hall, Oxford OX1 4AR, UK}

\author[0000-0003-0227-3451]{J.~P. Anderson}
\affiliation{European Southern Observatory, Alonso de C\'ordova 3107, Casilla 19, Santiago, Chile}
\affiliation{Millennium Institute of Astrophysics MAS, Nuncio Monsenor Sotero Sanz 100, Off. 104, Providencia, Santiago, Chile}

\author[0000-0002-5221-7557]{C.~Ashall}
\affiliation{Department of Physics, Virginia Tech, Blacksburg, VA 24061, USA}

\author[0000-0001-9629-5250]{W. Bate}
\affiliation{Astrophysics Research Centre, School of Mathematics and Physics, Queen's University Belfast, BT7 1NN, UK}

\author[0000-0002-5654-2744]{R. Fender}
\affil{Department of Physics, University of Oxford, Denys Wilkinson Building, Keble Road, Oxford OX1 3RH, UK}

\author[0000-0003-2375-2064]{C.~P. Guti\'errez}
\affiliation{Institut d’Estudis Espacials de Catalunya (IEEC), E-08034 Barcelona, Spain}
\affiliation{Institute of Space Sciences (ICE, CSIC), Campus UAB, Carrer de Can
Magrans, s/n, E-08193 Barcelona, Spain \\}

\author[0000-0003-4253-656X]{D.~A.~Howell}
\affiliation{Las Cumbres Observatory, 6740 Cortona Drive, Suite 102, Goleta, CA 93117-5575, USA}
\affiliation{University of California, Santa Barbara}

\author[0000-0003-1059-9603]{M. E. Huber}
\affil{Institute for Astronomy, University of Hawai'i, 2680 Woodlawn Drive, Honolulu, HI 96822, USA}

\author[0000-0002-3968-4409]{C. Inserra}
\affiliation{Cardiff Hub for Astrophysics Research and Technology, School of Physics \& Astronomy, Cardiff University, Queens Buildings, The Parade, Cardiff, CF24 3AA, UK\\}

\author{G. Leloudas}
\affiliation{DTU Space, National Space Institute, Technical University of Denmark, Elektrovej 327, DK-2800 Kgs. Lyngby, Denmark\\}

\author{L.~A.~G.~Monard}
\affiliation{Kleinkaroo Observatory, Calitzdorp, Western Cape
South Africa\\}

\author[0000-0003-3939-7167]{T.~E. M\"uller-Bravo}
\affiliation{Institute of Space Sciences (ICE, CSIC), Campus UAB, Carrer de Can
Magrans, s/n, E-08193 Barcelona, Spain \\}
\affiliation{Institut d’Estudis Espacials de Catalunya (IEEC), E-08034 Barcelona, Spain}

\author{B. J. Shappee}
\affiliation{Institute for Astronomy, University of Hawaii, 2680 Woodlawn Drive, Honolulu HI 96822, USA}

\author[0000-0001-9535-3199]{K. W. Smith}
\affil{Astrophysics Research Centre, School of Mathematics and Physics, Queen's University Belfast, BT7 1NN, UK}

\author[0000-0003-0794-5982]{G.~Terreran}
\affiliation{Las Cumbres Observatory, 6740 Cortona Drive, Suite 102, Goleta, CA 93117-5575, USA}

\author[0000-0003-2858-9657]{J. Tonry}
\affiliation{Institute for Astronomy, University of Hawaii, 2680 Woodlawn Drive, Honolulu HI 96822, USA}

\author{M. A. Tucker}
\affiliation{Department of Astronomy, The Ohio State University, 140 West 18th Avenue, Columbus, OH, USA}
\affiliation{Department of Physics, The Ohio State University, 191 West Woodruff Ave, Columbus, OH, USA}
\affiliation{Center for Cosmology and Astroparticle Physics, The Ohio State University, 191 West Woodruff Ave, Columbus, OH, USA}

\author[0000-0002-1229-2499]{D. R. Young}
\affil{Astrophysics Research Centre, School of Mathematics and Physics, Queen's University Belfast, BT7 1NN, UK}
 
\author[0000-0002-9085-8187]{A. Aamer}
\affiliation{Institute for Gravitational Wave Astronomy and School of Physics and Astronomy, University of Birmingham, Birmingham B15 2TT, UK\\}
\affiliation{Astrophysics Research Centre, School of Mathematics and Physics, Queen's University Belfast, BT7 1NN, UK}

\author[0000-0002-1066-6098]{T.-W. Chen}
\affiliation{Graduate Institute of Astronomy, National Central University, 300 Jhongda Road, 32001 Jhongli, Taiwan}

\author[0000-0003-2132-3610]{F. Ragosta}
\affiliation{INAF, Osservatorio Astronomico di Roma, via Frascati 33, I-00078 Monte Porzio Catone (RM), Italy}
\affiliation{Space Science Data Center – ASI, Via del Politecnico SNC, 00133 Roma, Italy}

\author[0000-0002-1296-6887]{L. Galbany}
\affiliation{Institute of Space Sciences (ICE, CSIC), Campus UAB, Carrer de Can
Magrans, s/n, E-08193 Barcelona, Spain \\}
\affiliation{Institut d’Estudis Espacials de Catalunya (IEEC), E-08034 Barcelona, Spain}

\author[0000-0002-1650-1518]{M.~Gromadzki}
\affiliation{Astronomical Observatory, University of Warsaw, Al. Ujazdowskie 4, 00-478 Warszawa, Poland}

\author[0000-0003-3393-9383]{L. Harvey}
\affiliation{School of Physics, Trinity College Dublin, The University of Dublin, Dublin 2, D02 PN40 Ireland}

\author{P.~Hoeflich}
\affiliation{Department of Physics, Florida State University, 77 Chieftan Way, Tallahassee, FL 32306, USA}

\author{C.~McCully}
\affiliation{Las Cumbres Observatory, 6740 Cortona Drive, Suite 102, Goleta, CA 93117-5575, USA}
\author[0000-0001-9570-0584]{M.~Newsome}
\affiliation{Las Cumbres Observatory, 6740 Cortona Drive, Suite 102, Goleta, CA 93117-5575, USA}
\affiliation{University of California, Santa Barbara}

\author{E.~P.~Gonzalez}
\affiliation{Las Cumbres Observatory, 6740 Cortona Drive, Suite 102, Goleta, CA 93117-5575, USA}
\affiliation{University of California, Santa Barbara}

\author{C.~Pellegrino}
\affiliation{Las Cumbres Observatory, 6740 Cortona Drive, Suite 102, Goleta, CA 93117-5575, USA}
\affiliation{University of California, Santa Barbara}

\author[0009-0009-2627-2884]{P. Ramsden}
\affiliation{School of Physics and Astronomy, University of Birmingham, Birmingham B15 2TT, UK}
\affiliation{Institute for Gravitational Wave Astronomy, University of Birmingham, Birmingham B15 2TT, UK}

\author[0000-0001-5654-0266]{M.~P\'erez-Torres}
\affiliation{Instituto de Astrof\'isica de Andaluc\'ia (IAA-CSIC),
Glorieta de la Astronom\'ia s/n, E-18008 Granada, Spain}
\affiliation{School of Sciences, European University Cyprus, Diogenes street, Engomi, 1516 Nicosia, Cyprus}

\author[0009-0008-2579-1810]{E. J. Ridley}
\affiliation{School of Physics and Astronomy, University of Birmingham, Edgbaston, B15 2TT, England}
\affil{Institute for Gravitational Wave Astronomy and School of Physics and Astronomy, University of Birmingham, Birmingham B15 2TT, UK\\}

\author[0000-0002-6527-1368]{X.~Sheng}
\affiliation{Astrophysics Research Centre, School of Mathematics and Physics, Queen's University Belfast, BT7 1NN, UK}

\author[0009-0002-9460-9900]{J.~Weston}
\affiliation{Astrophysics Research Centre, School of Mathematics and Physics, Queen's University Belfast, BT7 1NN, UK}

\begin{abstract}

We present multi-wavelength photometry and spectroscopy
of SN 2022jli, an unprecedented Type Ic supernova
discovered in the galaxy NGC\,157 at a distance of $\approx$ 23
Mpc. The multi-band light curves reveal many remarkable
characteristics. Peaking at a magnitude of $g=15.11\pm0.02$, the 
high-cadence photometry reveals 12.5$\pm0.2\,$day periodic undulations superimposed on the 200 day supernova decline. This periodicity is
observed in the light curves from nine separate filter and instrument
configurations with peak-to-peak amplitudes of $\simeq$ 0.1 mag. This is the first
time that repeated periodic oscillations, over many cycles, have been
detected in a supernova light curve. SN 2022jli also displays an
extreme early excess which fades over $\approx$ 25 days followed by a rise to a
peak luminosity of $L_{\rm opt} = 10^{42.1}$\,erg\,s$^{-1}$. Although the exact
explosion epoch is not constrained by data, the time from explosion to
maximum light is $\gtrsim$ 59 days. The luminosity can
be explained by a large ejecta mass ($M_{\rm
ej}\approx12\pm6$M$_{\odot}$) powered by $^{56}$Ni but we find
difficulty in quantitatively modelling the early excess with
circumstellar interaction and cooling. Collision between the supernova
ejecta and a binary companion is a possible source of this emission.
We discuss the origin of the periodic variability in the
light curve, including interaction of the SN ejecta with nested shells
of circumstellar matter and neutron stars colliding with binary
companions.

\end{abstract}


\section{Introduction} \label{sec:intro}

Stars with zero age main sequence masses (M$_{\rm ZAMS}$) greater than 8 M$_\odot$ end their lives as core-collapse supernovae \citep[CCSNe;][]{2009ARA&A..47...63S,2012ARA&A..50..107L}, producing a diverse range of transients \citep[e.g.][]{2017hsn..book..195G,2019NatAs...3..717M}. The variety in the observable properties of these SNe is thought to be dependent on the initial mass, metallicity, binarity and mass-loss history of the progenitor star.
Hydrogen-poor CCSNe are referred to as stripped-envelope (SE)SNe due to significant mass loss of the progenitor, removing hydrogen, and in some cases helium, from the stellar envelope. 
SESNe classified as Type Ic do not show hydrogen or helium in their optical spectra, although the extent of helium-stripping is still uncertain \citep{2012MNRAS.422...70H,2021ApJ...908..150W}.
Envelope stripping can occur through strong stellar line-driven winds 
\citep[e.g.][]{2005A&A...442..587V,2020A&A...634A..79S} 
or interaction with a binary companion \citep{1992ApJ...391..246P}.

Evidence for periodicity has been searched for in supernova light curves. \citet{2016ApJ...826...39N} investigated the undulations in the superluminous SN 2015bn but were limited by the duration of their time series and could not reliably identify periodicity. \citet{2023A&A...670A...7W}  suggested a 32\,$\pm\,$6 day repeating pattern in the declining light curve of SN 2020qlb (an explosion somewhat similar to SN 2015bn). However, insufficient cycles were observed to perform robust statistical checks for periodicity. 
 \citet{2015AJ....149....9M} and \citet{2013MNRAS.433.1312F} suggested a periodicity in the optical light curve of SN 2009ip, but \citet{2015MNRAS.453.3886F} subsequently found no evidence for the periodicity in  extensive $R$-band data. The light curve of the luminous, fast optical transient AT 2018cow was subject to periodicity searches and while none were found in the optical, marginal evidence for periodicity in the variable X-ray light curve was suggested \citep{2018MNRAS.480L.146R,2019MNRAS.487.2505K,2019ApJ...872...18M}. 
Perhaps the most promising detection of periodicity in supernova emission is in the  radio light curves of SN 1979C \citep{1992ApJ...399..672W} and SN 2001ig \citep{2004MNRAS.349.1093R}, which have been  attributed to fluctuations in the density of the circumstellar medium (CSM) produced by binary stellar wind interactions.

In this paper we present an extensive follow-up campaign of the Type Ic SN 2022jli from $\sim$ $-50$ days to +200 days relative to maximum light.  SN 2022jli presents an unusually long-lived, luminous early excess followed by a 
long rise time, and slow spectroscopic evolution. The extensive, almost daily, photometric coverage of this bright SN for 200 days after peak indicates a  
periodic variability  ($P= 12.5 \pm 0.2$-day) observed in multiple bands and instruments, with amplitude of  order $1\%$ of the peak bolometric luminosity of the SN. 

\section{Discovery and Classification}\label{sec:disc}

Libert Monard discovered a transient in NGC 157 from Kleinkaroo Observatory and submitted the discovery report  on the Transient Name Server (TNS) as AT 2022jli on 2022 May 5.17 UT at an unfiltered magnitude $\simeq14$\,mag\,\citep{2022TNSTR1198....1M}. With the ATLAS survey \citep{2018PASP..130f4505T,2020PASP..132h5002S}, we independently detected the object (internal name ATLAS22oat) on 2022 May 16.41 (at $o$=14.3\,mag). 
The original TNS Discovery Report of \cite{2022TNSTR1198....1M} registered the object with an astrometric  error of 14\arcsec\footnote{L. Monard  corrected this a day later in the TNS Comment for AT 2022jli, but the TNS database coordinates remained in error.} and our ATLAS transient server \citep{2020PASP..132h5002S}, which dynamically links to TNS discoveries, did
not associate the two sources. The ATLAS  automated TNS registration triggered a new source ($\alpha=8.69038$\degree\, $\delta=-8.38668$\degree), a discovery report, and name (AT 2022jzy). To prevent confusion, AT 2022jzy was manually removed entirely from the TNS records and the original incorrect coordinates of AT2022jli were replaced with those from ATLAS while preserving L. Monard's discovery credit (O. Yaron, priv. comm.). A low resolution 
($R=100$) spectrum from a 0.35m telescope \citep{2022TNSCR1261....1G} indicated a likely type Ic. This classification was confirmed  \citep{2022TNSAN.116....1C} by
the extended Public ESO Spectroscopic Survey of Transient Objects \citep[ePESSTO+;][]{2015A&A...579A..40S} survey. 

The SN is offset by $35\farcs2$N and $15\farcs88$W from the centre of its host galaxy NGC 157, which has a redshift of $z = 0.0055$. The kinematic distance on the NASA/IPAC Extragalactic Database (NED), from the recessional velocity (corrected for Virgo infall and assuming $H_0=73$\kmsMpc) is $D = 23\pm2$\,Mpc. Distance estimates from the Tully-Fisher and Sosies methods \citep[e.g.][]{2002A&A...393...57T,2013AJ....146...86T} have a large range from 11 to 29\,Mpc and we adopt the kinematic distance $D = 23\pm2$\,Mpc throughout the rest of this paper.  The foreground Milky Way reddening is  $A_V= 0.1186$ \citep{2011ApJ...737..103S} and given the position of SN 2022jli, some internal host extinction is likely present. We do not account for possible host extinction, but this is likely to be low due to the lack of narrow Na I D absorption in the spectra \citep{2012MNRAS.426.1465P} and our main results are not sensitive to this choice. The spectroscopy and photometry presented in this paper have been corrected for foreground galactic extinction and the spectra have been shifted into the rest-frame. We note that NGC 157 also hosted the Type Ic SN 2009em \citep{2009CBET.1798....1M}.

\section{Observations}

\subsection{Imaging and Photometry}
The first observations of SN 2022jli were reported to the TNS by L. Monard \citep{2022TNSTR1198....1M}. To our knowledge there are no pre-explosion non-detections available as the object had just emerged from solar conjunction. Monard reported four epochs of unfiltered CCD photometry from observations taken at the Kleinkaroo Observatory between 2022 May 5 (MJD 59704) to 2022 May 22 UT (MJD 59721).

The Asteroid Terrestrial-impact Last Alert System \citep[ATLAS;][]{2018PASP..130f4505T} began observing at the position of SN 2022jli  on 2022 May 16 (MJD 59715) in normal survey operations. ATLAS is a quadruple 0.5\,m telescope system using broad orange (\textit{o}, $5600-8200~\rm \AA$) and cyan (\textit{c}, $4200-6500~\rm\AA$) filters. The combined four-telescope system surveys the observable sky to a typical 5$\sigma$ depth of $\sim$ 19 mag and a cadence of 1-2 days. ATLAS photometry and astrometry are calibrated with the all-sky reference catalog sky \citep[refcat2;][]{2018ApJ...867..105T}. ATLAS photometry for SN 2022jli were obtained by forcing photometry at the location using the ATLAS forced photometry server \citep{2021TNSAN...7....1S} and adopting a 3$\sigma$ clipped nightly mean.

The Zwicky Transient Facility \citep[ZTF;][]{2019PASP..131a8002B} observed the field beginning on 2022 July 03 UT (MJD 59763), giving the object the internal name is ZTF22aapubuy. ZTF photometry in both \textit{g} and  \textit{r} bands was obtained from the ZTF public stream using the Lasair\footnote{https://lasair-ztf.lsst.ac.uk/object/ZTF22aapubuy/} broker \citep{2019RNAAS...3a..26S}.

Photometry from ASAS-SN \citep{2014ApJ...788...48S} beginning on 2022 May 09 UT (MJD 59708) was obtained using the ASAS-SN Sky Patrol website \footnote{https://asas-sn.osu.edu} \citep{2017PASP..129j4502K}. We adopt a cut-off MJD of 59762, after which we do not include ASAS-SN \textit{g}-band photometry in favour of higher signal-to-noise ZTF \textit{g}-band.

We triggered follow-up photometric observations of SN 2022jli using the IO:O camera at the 2m Liverpool Telescope \citep[LT;][]{2004SPIE.5489..679S}. Using the LT we obtained six epochs of $ugriz-$band observations and an additional $griz-$band observation between MJD 59817-59894.

\textit{griBV}-band photometry was obtained through the Global Supernova Project using the 1m Las Cumbres Observatory \citep[LCO;][]{2013PASP..125.1031B}. Additional V-band observations were recovered from acquisition images taken with the ESO Faint Object Spectrograph and Camera (v.2) \citep[EFOSC2;][]{2008Msngr.132...18S} on the ESO 3.58m New Technology Telescope \citep[NTT;][]{1983ESOC...17..173W} during spectroscopic follow-up by PESSTO \citep{2015A&A...579A..40S}.

All CCD reductions were performed using instrument specific pipelines. Photometric measurements for the LCO-1m, EFOSC2, and \textit{griz} LT-IO:O data were made using \texttt{AUTOPHOT} \citep{2022A&A...667A..62B} without host subtraction. Photometry in \textit{griz}-bands were calibrated against Pan-STARRS field stars \citep{2020ApJS..251....7F} and \textit{BV}-band photometry was calibrated using the APASS catalog \citep{2016yCat.2336....0H}. LT $u-$band measurements were performed using the PSF package\footnote{http://github.com/mnicholl/photometry-sans-frustration} \citep{2023arXiv230702556N} and calibrated against the Sloan Digital Sky Survey catalog \citep[SDSS;][]{2015ApJS..219...12A}.

The \textit{Gaia} \citep{2016A&A...595A...1G} satellite, operated by the European Space Agency (ESA), observed SN 2022jli (internal name Gaia22cbu) between 2022 May 11 (MJD 59710) and 2022 June 30 UT (MJD 59760). The \textit{Gaia} Science Alerts Project \citep{2021A&A...652A..76H} reported three epochs of $G-$band photometry\footnote{http://gsaweb.ast.cam.ac.uk/alerts/alert/Gaia22cbu/}. We assume a pessimistic \textit{Gaia} photometric uncertainty of 0.1 mag.

\begin{figure*}
    \centering
    \includegraphics[width=\linewidth]{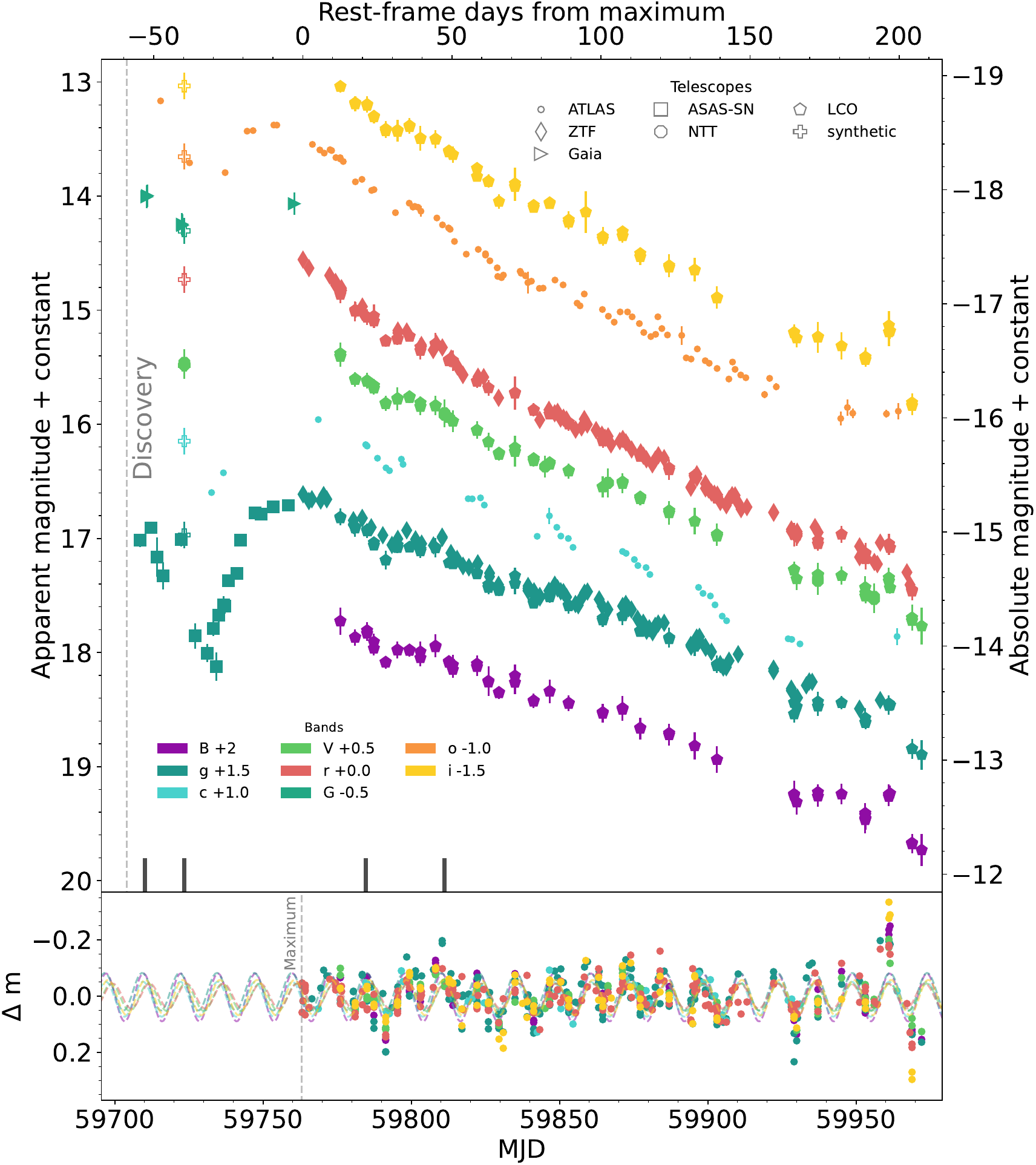}

    \caption{Top: Multi-colour extinction-corrected light curves of SN 2022jli including photometric errors. Black lines along the bottom of the plot indicate spectroscopic observations. Unfilled points are synthetic photometry performed on the EFOSC2 spectrum. Bottom: Light curves after de-trending with a second-degree polynomial fit including only the region after maximum-light (vertical dashed line). Over-plotted for each band are the best fit sinusoids produced in the GLS analysis (Section \ref{sec:GLS}). We retain the colors from the top panel and remove variation of point shape for visual clarity, including only observations after MJD 59750.}
    \label{fig:phot}
\end{figure*}

Ultraviolet (UV) and optical photometry of SN 2022jli was performed with the Ultra-Violet and Optical Telescope \citep[UVOT;][]{2005SSRv..120...95R} onboard the \textit{Neil Gehrels Swift Observatory} \citep[\textit{Swift};][]{2004ApJ...611.1005G}. \textit{Swift} observed the field 13 times between 2022 August 17 (MJD 59808) and 2022 December 27 (MJD 59940) in the $U$, $B$, $V$, $UVW1$, $UVM2$ and $UVW2$ bands. The images in each filter were co-added, and SN magnitudes were extracted using standard tasks within the \texttt{HEASOFT}\footnote{
https://heasarc.gsfc.nasa.gov/lheasoft} package. A small aperture of $3\arcsec$ was chosen, and an aperture correction was applied, following \cite{2009AJ....137.4517B}. Without template subtraction most $UVW1$, $UVM2$, and $UVW2$ exposures were non-detections. Keeping only detections greater than the  limiting magnitude we retain only one epoch of $UVM2$ photometry but retain most observation in the \textit{UBV}-bands. 
The extinction-corrected light curve of SN2022jli is shown in Figure \ref{fig:phot}.

\subsection{Radio Observation}

We obtained a single radio observation at the position of SN 2022jli on 2022 Sep 19, starting at 22:15\,UT with the \textit{enhanced}-Multi Element Radio Linked Interferometer Network (e-Merlin) (DD14001, PI: Rhodes). Observations were obtained at a central frequency of 5.08\,GHz with a bandwidth of 512\,MHz. The observation consisted of 6\,minute scans of the target interleaved with 2\,minute scans on the phase calibrator (J0039-0942). The observation ended with a scan of the flux calibrator (J1331+3030) and the bandpass calibrator (J1407+2827). The data were processed using the \textit{e}-MERLIN custom \textsc{casa}-based pipeline \citep[Version 5.8, ][]{2021ascl.soft09006M}.

The pipeline averages the data in both time and frequency space, flags the data for radio frequency interference, performs bandpass and complex gain calibration and splits out the calibrated target field. We performed some further flagging and imaged the data within \textsc{casa}. During the observation two of the six antennas dropped out which impacted the quality of the final image. We did not detect any radio emission at the position of SN 2022jli and measure a final rms noise of about 52\,$\mu$Jy/beam (and a 3$\sigma$ upper limit of 156\,$\mu$Jy/beam). 

\subsection{Spectroscopy} \label{sec:spectroscopy}

We present our spectra spanning three epochs, $-39$ days to +47 days with respect to $g-$band maximum and also show the low resolution spectrum of \cite{2022TNSCR1261....1G} from the TNS (additional spectra will be presented in a separate publication). Foreground Galactic reddening was corrected using the \texttt{dust\_extinction} package of  \texttt{Astropy} following the \citet{1999PASP..111...63F} reddening law. All spectra presented in this work will be made available via the WISeREP repository \citep{2012PASP..124..668Y}.

Follow-up spectroscopy was acquired using the EFOSC2 at the 3.58m NTT  \citep{2008Msngr.132...18S}, at two epochs through ePESSTO+. The first EFOSC2 spectrum was taken on 2022 May 24.42 UT (MJD = 59723.42), and our final spectrum on MJD = 59811.15.  The EFOSC2 grism used for the spectral sequence was  Gr\#13 (3685 - 9315 \AA).
Data reductions were performed using the PESSTO pipeline, which includes flat-fielding, bias-subtraction, wavelength and telluric correction, and flux calibration as described by \cite{2015A&A...579A..40S}.

One epoch from the University of Hawaii 2.2\,m telescope was obtained on 2022 July 24.62 (MJD 59784.62) using SNIFS \citep{2004SPIE.5249..146L}. The SNIFS spectrum was reduced using the Spectroscopic Classification of Astronomical Transients (SCAT) Survey pipeline \citep{2022PASP..134l4502T}.

\begin{figure*}

    \includegraphics[width=\linewidth]{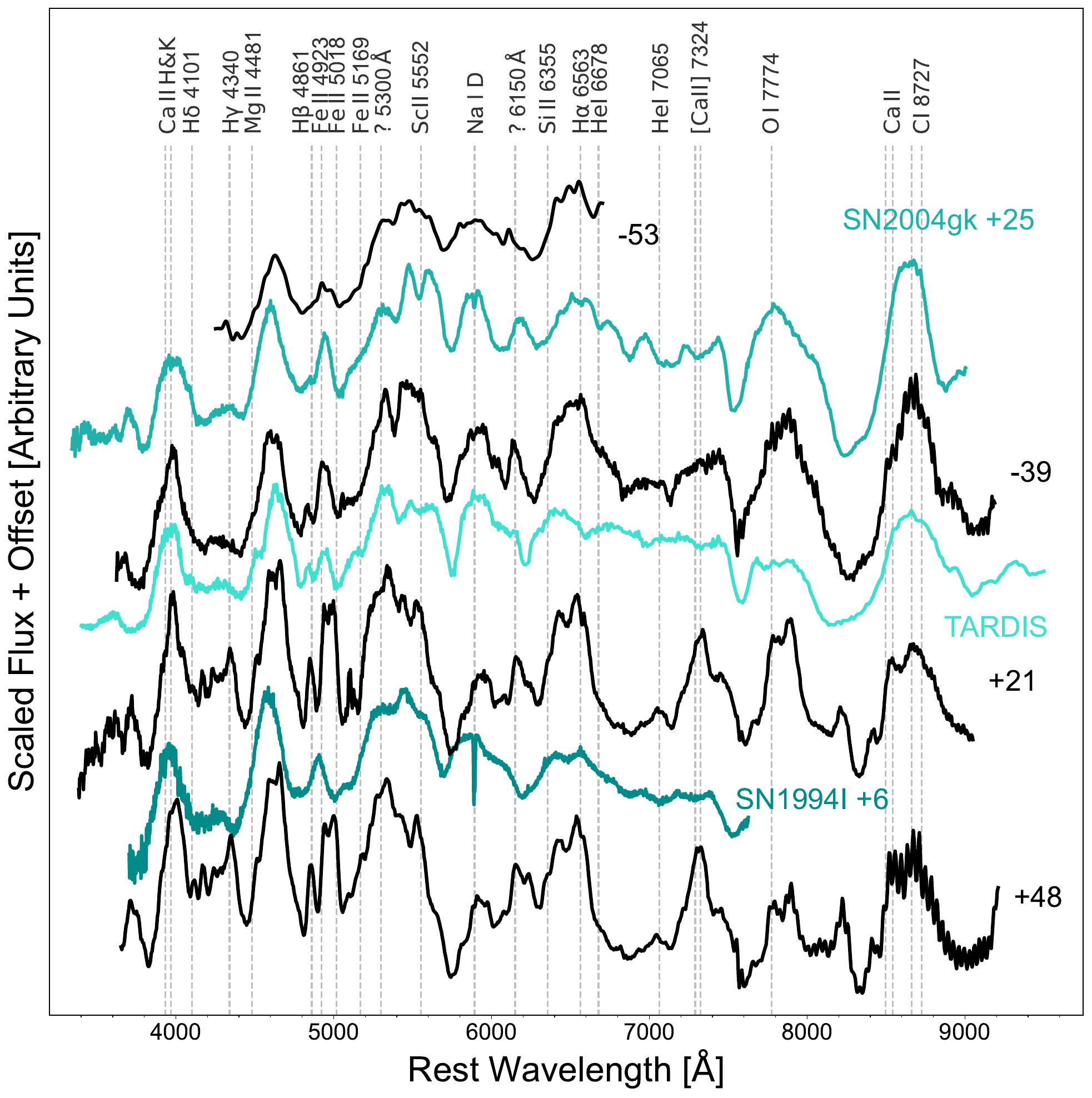}
    \caption{Spectroscopic follow-up observation of SN 2022jli. Spectra are labelled by time since maximum light in the rest frame. Common supernova lines are marked as vertical lines corresponding to the rest wavelengths. We include here the \citet{2022TNSCR1261....1G} classification spectrum. A TARDIS \citep{2014MNRAS.440..387K} model of H, He free material is also included. Spectra for the Type Ic SNe SN2004gk \citep{2019MNRAS.482.1545S} and SN1994I \citep{2014AJ....147...99M} were obtained from WISeREP \citep{2012PASP..124..668Y}.
    }
    \label{fig:spectra}
\end{figure*}

The extinction corrected spectra are shown in Figure \ref{fig:spectra}.

\section{Analysis}

\subsection{Light Curve}\label{subsec:lc}

SN 2022jli fades for $\sim$ 25 days after discovery in the \textit{go}-bands, using a linear fit we measure a decline rate of $\sim$ 5 mag (100 d)$^{-1}$, which is incompatible with \Co\,decay. This early decline is not well sampled with low-cadence coverage from only ASAS-SN ($g$), ATLAS ($o$), \textit{Gaia}, and a single $V$-band observation from the NTT. The light curve begins to rise after MJD 59734, reaching maximum light on MJD 59763, indicating the main peak is at $\sim$ 59 rest-frame days from discovery. SN 2022jli exhibits a significantly longer rise to maximum light than literature samples of SESNe \citep{2019MNRAS.485.1559P}, comparable to a small subset of slowly evolving Ibc SNe \citep[e.g.][]{2005ApJ...631L.125A,2016MNRAS.457..328L,2016A&A...592A..89T,2018A&A...609A.106T,2022arXiv221009402K}. The SN eventually fades in the optical at a rate of $\sim$ 1 mag (100 d)$^{-1}$ which is compatible with with \Co\,decay suggesting a radioactively-powered main peak \citep{1989ApJ...346..395W}. The $r$-band declines 0.4 mag (100 d)$^{-1}$ faster than the $g$-band, showing an unusual evolution towards bluer colors in $g-r$ over time. The double-peaked early light curve indicates that \Ni\ decay cannot account for the full structure of the light curve. 

As the light curve slowly fades from peak, the extensive high-cadence photometry
of SN 2022jli capture clear undulations in the photometry.  The undulations are visible in multiple bands
($B$ to $i$) and across different telescope and instrument combinations, indicating that this is 
neither an instrumental or calibration effect. We subtract the SN continuum and reveal these undulations more clearly in the bottom panel of Figure \ref{fig:phot}. We identify repeating bumps with a consistent timescale for all filters and discuss this
in detail in Section \ref{sec:GLS}.

During the initial decline we observe an increase in the ASAS-SN $g-$band photometry on MJD 59722.37. Seeking to confirm the validity of this observation, we perform synthetic photometry on the EFOSC2 spectrum taken on MJD 59723.42, which was calibrated to the $V$-band acquisition image. We include the synthetic photometry on Figure \ref{fig:phot}, the errors on the points are consistent with the ASAS-SN $g-$band photometry and other contemporary photometric observations in $G$ and $o$-band. We interpret this  epoch as a short lived, luminous episode during the initial excess.

\subsubsection{Bolometric Light Curve} \label{subsec:bolo}

We compute a pseudo-bolometric light curve by integrating under the $BgcVrGoiz-$band observations using the publicly available code \texttt{SUPERBOL} \citep{2018RNAAS...2..230N}.
From our bolormetric light curve we measure a peak luminosity $L_{\rm opt} = 10^{42.08 \pm 0.04}$ erg\,s$^{-1}$,  which is within the typical range of SESNe found by \cite{2019MNRAS.485.1559P}.  The total integrated luminosity (across the wavelength range covered by our filters) 
is $E_{\rm opt}$ $\approx 2.5 \times 10^{49}$ ergs.

We compare the (pseudo-)bolometric light curve to other SNe including  normal SESNe and those with double-peaked light curves in Figure \ref{fig:compare}. SN 2022jli exceeds the peak brightness of the Type Ic SN 2007gr \citep{2009A&A...508..371H}, and Ib SNe  SN 2008D \citep{2008Natur.453..469S,2009ApJ...702..226M} and the relatively faint Type Ib SN 2007Y \citep{2009ApJ...696..713S}, showing a significantly more luminous broad peak and slower decay. SN 2007gr and SN 2007Y both display a monotonic rise and smooth decline, typical of normal SN Ibc, unlike the double-peaked structured light curve of SN 2022jli. The overall shape of the light curve resembles the unusual Type Ic iPTF15dtg \citep{2016A&A...592A..89T}. Both SNe have a fast declining early excess with a broad persistent maximum.
SN 2005bf \citep{2005ApJ...631L.125A} has an early peak and broad maximum but is significantly more luminous than SN 2022jli and declines significantly faster.

\subsubsection{Light Curve Modelling}
\label{sec:lightcurve_model}

We model the light curve using simple models to derive a representative ejecta-mass estimate for SN 2022jli using the Modular Open Source Fitter for Transients \citep[MOSFiT;][]{2018ApJS..236....6G}. MOSFiT is a publicly available code which we use to fit semi-analytic models to the multiband observed light curves of SN 2022jli. We use two models, one where we model only the broad ‘main’ peak assuming radioactive decay of \Ni\ as the only energy source \citep{1982ApJ...253..785A,1994ApJS...92..527N}, and another where we fit the full light curve interpreting the initial excess as shock cooling emission (SCE) from interaction with a CSM and a subsequent radioactively powered ‘main’ peak \citep{2013ApJ...773...76C}. All model fitting was performed using the dynamic nested sampler \texttt{DYNESTY} package \citep{2020MNRAS.493.3132S} option in MOSFiT.

\begin{figure}
    \centering
    \includegraphics[width=1\linewidth]{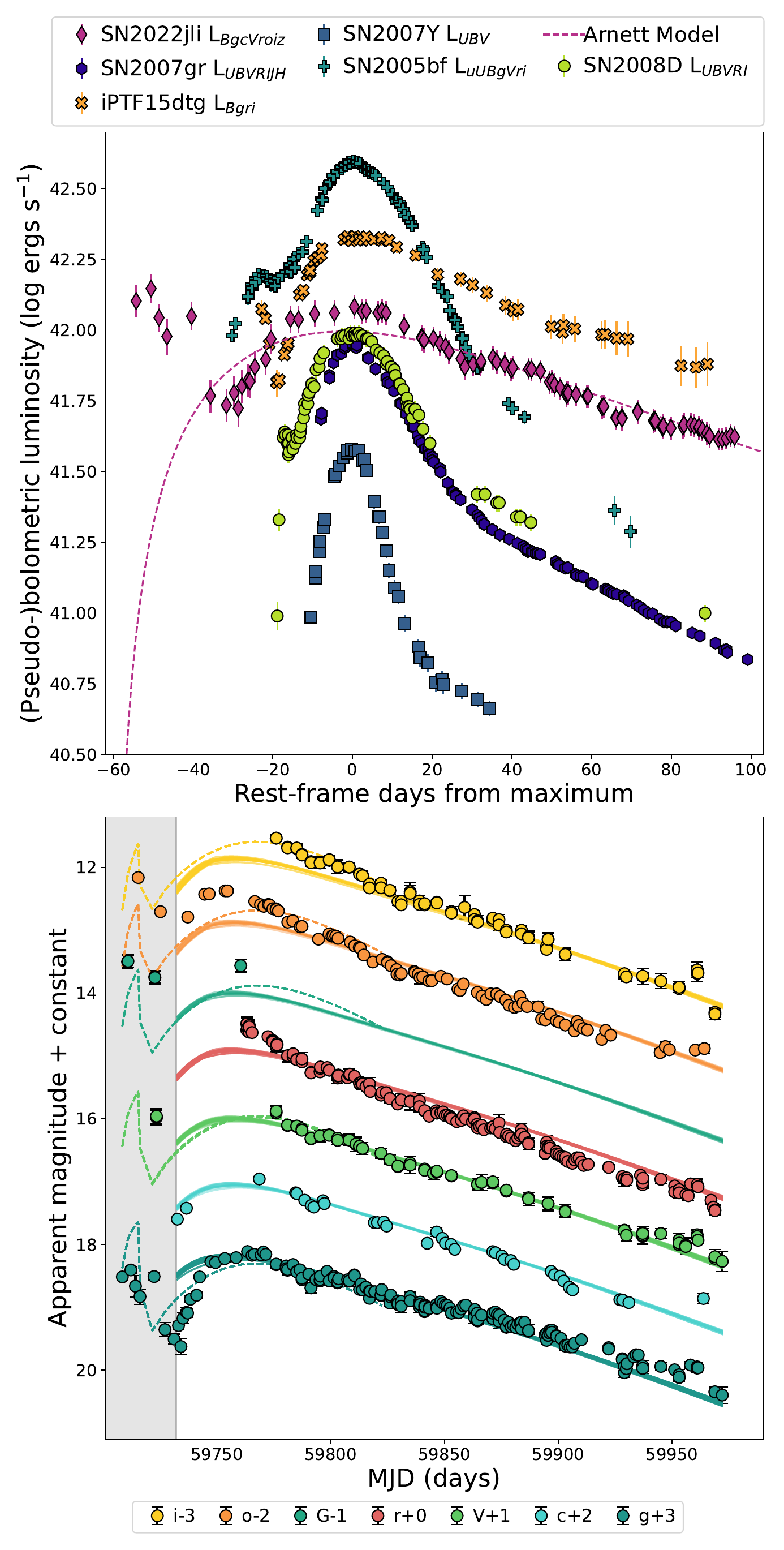}
    \caption{Top: Pseudo-bolometric light curve comparison of SN 2022jli and other SNe with prominent bumps and well studied Ibc SNe. The bolometric light curves of the Type Ic SN 2007gr \citep{2009A&A...508..371H}, Type Ib SN 2007Y \citep{2009ApJ...696..713S}, Type Ic iPTF15dtg \citep{2016A&A...592A..89T}, Type Ib
    SN 2008D \citep{2008Natur.453..469S,2009ApJ...702..226M} (pseudo-bolometric lightcurve calculated in \citet{2015MNRAS.452.3869N}), and Type Ib/c
    SN 2005bf \citep{2005ApJ...631L.125A} were also constructed using \texttt{SUPERBOL}. The dashed line represents an Arnett \citep{1982ApJ...253..785A} model fit to SN 2022jli light curve after MJD 59732. Bottom: MOSFiT \citep{2018ApJS..236....6G} multiband fit results for a \Ni\  explosion model (solid lines). We include the best scoring realization from a  CSM + \Ni\ model fit to the first 110 days of SN 2022jli as a dashed line.}
    \label{fig:compare}
\end{figure}

The lower panel in Figure \ref{fig:compare} shows the \Ni-only model fit to SN 2022jli. To construct this model we used the nickel driven explosion model built into MOSFiT \citep{1994ApJS...92..527N}, omitting any data during the early excess (before MJD 59732). We modify the priors of the model to require the explosion time to be before discovery i.e. MJD$_{\text{explosion}} <$ MJD$_{\text{discovery}}$. The opacity was fixed at $\kappa$=0.1 cm$^2$ g$^{-1}$ and $\kappa_{\gamma}=0.027$ cm$^2$ g$^{-1}$. With no prior on ejecta velocity the data require $v_{\rm ej} \approx 2500$\,km\,s$^{-1}$ and $M_{\rm ej}\approx 4\,$ M$_\odot$, this velocity is much lower than the  measurements of the Fe II lines  in the spectra (see Section \ref{sec:spec}). The posterior distribution for this fit is included in the Appendix (Figure \ref{fig:corner}). 

This model reproduces the maximum luminosity for the $g$ and $c$-bands but fits poorly to the redder-bands, underestimating the flux particularly in the $o$-band and $i$-bands. This simple model also fails to reproduce the fast $g$-band rise after the light curve dip and the $o$-band peak luminosity.%, perhaps this delayed rise is due to\Ni\ deposition deep within the core which is not taken into account in this simple model. 
Color differences between the model light curve and the observed data are likely due to the black body assumption made by MOSFiT, as the true spectrum is dominated by strong emission and absorption lines by the time of maximum light. Further detailed modelling is warranted using more sophisticated techniques, which is beyond the scope of this work. For comparison, we perform an additional Arnett model \citep{1982ApJ...253..785A} fit to the bolometric light curve with a $\chi^{2}$ fitting approach. Fixing $\kappa$=0.1 cm$^2$ g$^{-1}$ and $v_{\rm ej} \approx 3000$\,km\,s$^{-1}$ we require $M_{\rm ej}\approx 6$\,M$_\odot$, which is in agreement with the results from MOSFiT. This fit is included in Figure \ref{fig:compare}.

To gauge the systematic modelling uncertainties  within MOSFiT, we run the same nickel-driven model for the main peak with a range of $v_{\rm ej}$ to determine a probable range of $M_{\rm ej}$. A fixed ejecta velocity of $v_{\rm ej} = 3500$\,km\,s$^{-1}$ requires $M_{\rm ej}\approx 7\,$ M$_\odot$, $v_{\rm ej} = 6000$\,km\,s$^{-1}$ requires $M_{\rm ej}\approx 18$ M$_\odot$, and $v_{\rm ej} = 7000$\,km\,s$^{-1}$ forces $M_{\rm ej}\approx 21\,$ M$_\odot$. With no pre-explosion non-detections available to constrain the explosion time and large systematic errors on the model, we adopt an indicative mass range for SN 2022jli of $M_{\rm ej}\approx 12\pm6$M$_\odot$.

In the second scenario we consider the contribution of SCE following shock breakout of the ejecta through a dense CSM using the \Ni\ + CSM (\texttt{CSMNI}) model \citep{2013ApJ...773...76C}. SCE is a natural interpretation for a fast-declining excess shortly after explosion and is now regularly detected for a range of SN sub-types. The same assumptions are made for the opacities as before, but the explosion time is left as a free parameter of the fit. We modify the code so the interaction begins when the ejected material reach the inner radius of the CSM at $R_0$. Our results indicate a CSM radius $R_0 \sim 1 \text{AU}$, $M_{\rm ej}\,\sim 20$\,M$_\odot$, $M_{\rm CSM}\,\sim 26$\,M$_\odot$, and $f_{Ni}\sim0.01$ for the early excess to be powered by CSM interaction. Our model (fit to the first 110 days) results in poor match to the observed data, particularly the colors of the early excess and require physically improbable parameters, this model is shown in Figure \ref{fig:compare}.

Although the \citet{2013ApJ...773...76C} model implemented in MOSFiT is relatively simple, the very large ejecta masses required imply that this scenario is
physically unlikely. We return to this point in Section\,\ref{subsec:initalmax}. We also emphasise that there is no robust measurement of explosion time to constrain the model since the earliest epoch from the Monard observations are after the SN appeared from solar conjunction. 

A long rise time (exceeding 30 days) is a rare occurrence for Ibc SNe \citep{2016MNRAS.457..328L}: long duration light curves with rise times similar to SN 2022jli arise from only $\sim 6 - 10 \%$ of SNe Ibc in a bias corrected sample \citep{2022arXiv221009402K}. We estimate M$_{\rm ej}$ $=$ 12$\pm6$M$_{\odot}$ is required to provide the long rise to maximum light. An ejecta mass this extreme is rare and points to a high-mass progenitor star.

\subsection{Periodic Variability}

\label{sec:GLS}

\begin{figure}
    \centering
    \includegraphics[width=1\linewidth]{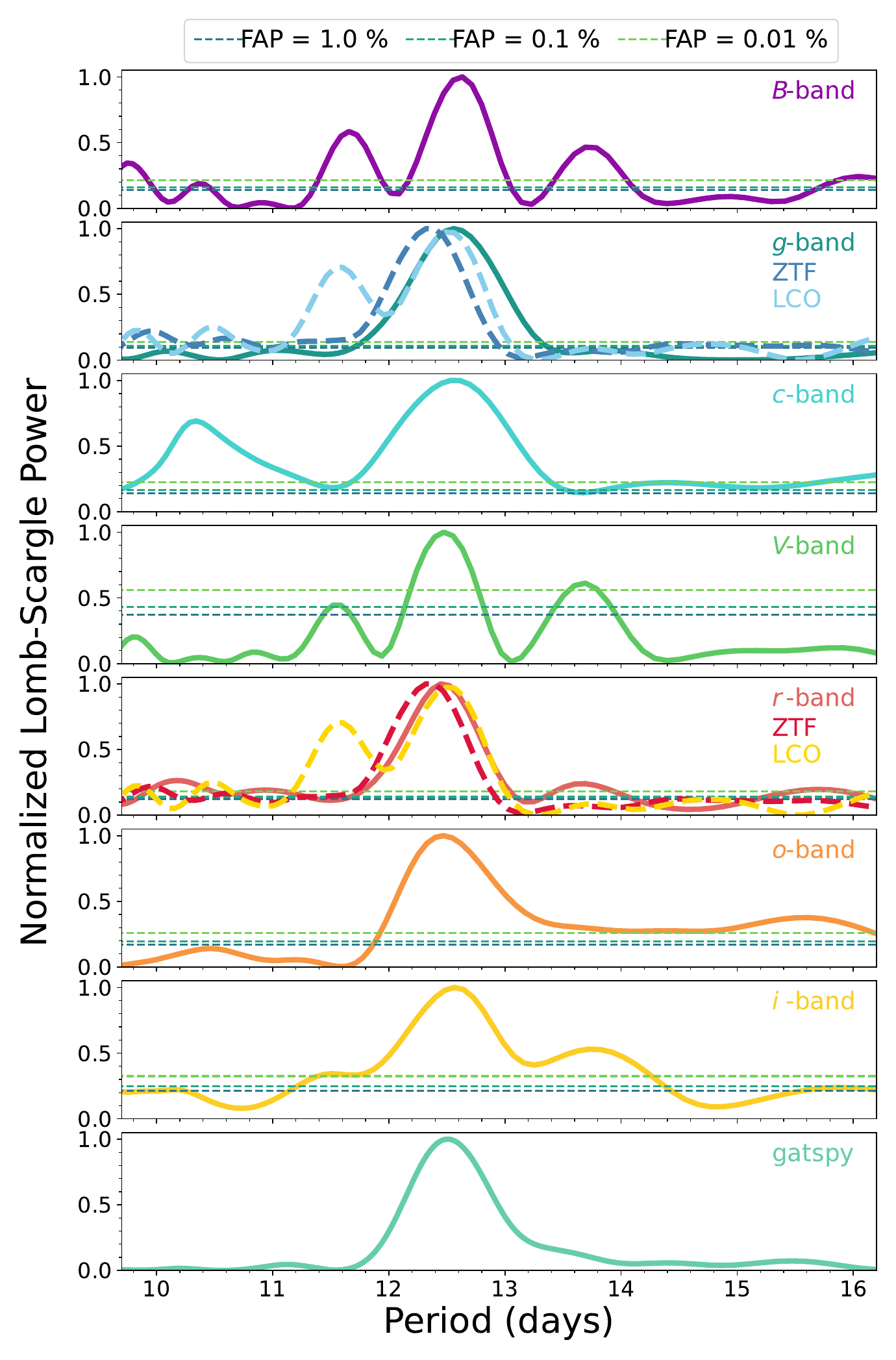}
    \includegraphics[width=1\linewidth]{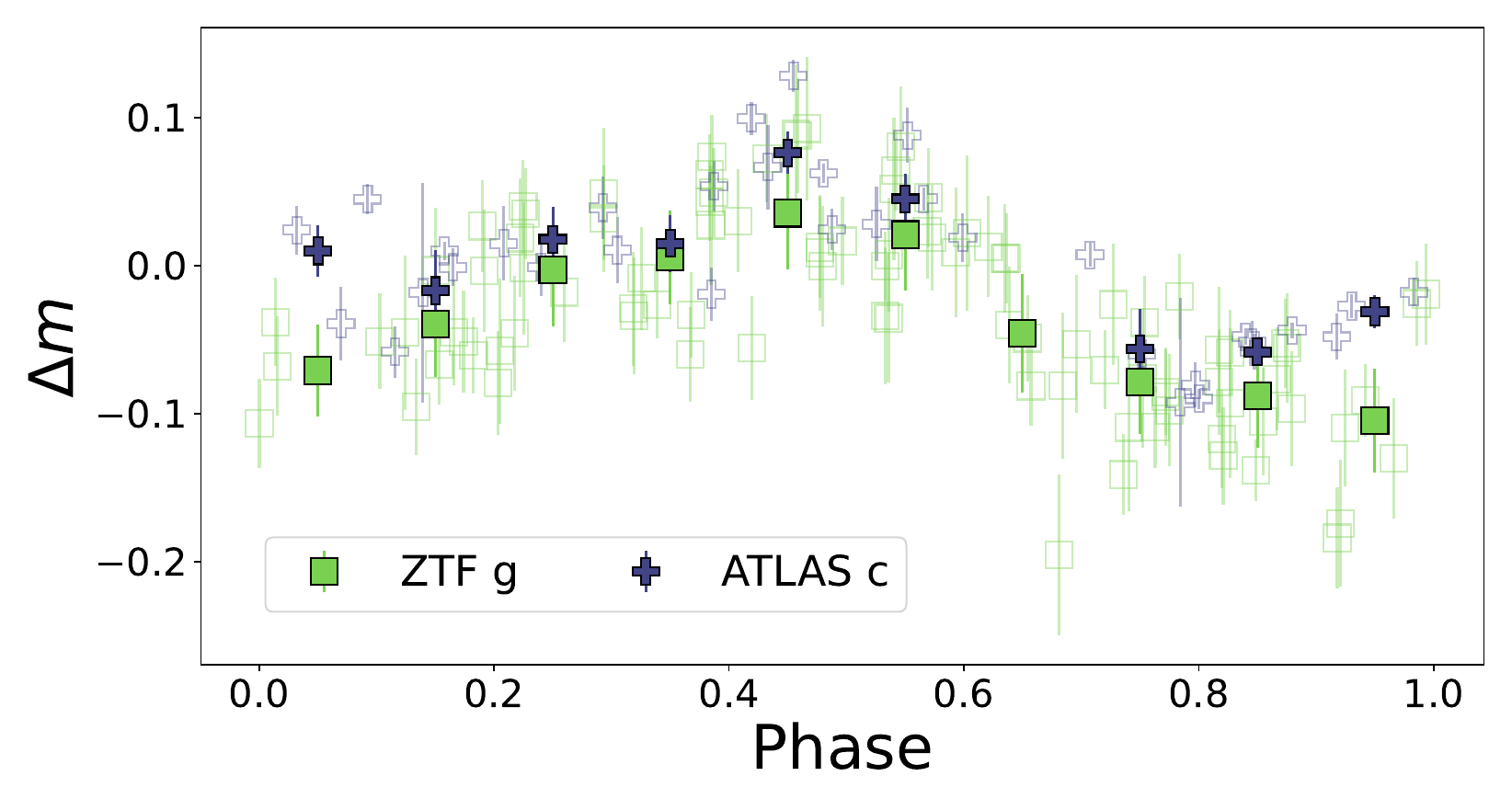}
    
    \caption{Top: Generalised Lomb-Scargle periodograms from  $BgcVroi$-bands where dashed horizontal lines correspond to the 1 \%, 0.1\% and 0.01\% FAP levels. The bottom panel shows a Lomb-Scargle periodogram generated from all bands simultaneously using \texttt{gatspy}. For the $gr$-bands additional periodograms are computed using only LCO and ZTF photometry, the associated FAP levels for these periodograms are not plotted. All periodograms regardless of band or telescope-instrument combination show a significant power at $\sim$12.5 days. 
    Bottom: Binned Phase-folded ZTF $g$-band and ATLAS $c$-band light curve. Each band has been folded on their respective best periods, the unbinned phase folded points are shown as unfilled points. }
    \label{fig:GLS}
\end{figure}

The declining light curve (Figure \ref{fig:phot}) shows $\sim 0.05$\,mag undulations, which are present across all bands and appear to repeat with a regular amplitude and period.  To search for and quantify any periodicity, we first removed
the decline signature from the light curve. From the post-peak light curve data (MJD $>$ 59760), we produced a residual light curve in each band by fitting and subtracting a fourth-order polynomial fit (the lowest order which removes the SN decline) between MJD 59760 and the end of the time series for each band. We applied this method to each of the \textit{BgcVroi}-bands independently and include the results in the lower panel of Figure \ref{fig:phot}.  The residual light curves show consistent oscillations over time across all bands. 
The periodicity in each of the residual light curves  was quantified by 
computing a periodogram using a Generalised Lomb-Scargle (GLS) method \citep{2009A&A...496..577Z}. The periodograms for each band and a phase folded light curve are shown in Figure \ref{fig:GLS}.

 We find the undulations have a dominant frequency of $\sim$ 0.08 days$^{-1}$ (or a period of $\sim12.5$ days), where significant power is observed across the period range of $12–13$~days, which is safely below the $\Delta t / 3$ cutoff adopted by \citet{2015AJ....149....9M} and \citet{2016ApJ...826...39N}, where $\Delta t$ is the length of the time series. The maximum GLS power in this region exceeds the 0.01\% false-alarm probability \citep[FAP; see][]{2009A&A...496..577Z} level in all bands (shown in Figure~\ref{fig:GLS}). The de-trended data reveal peaks and troughs with amplitudes in the range of $0.04 – 0.08$~mag across all bands. Motivated by the synchronized behavior of the multiband photometry, we compute a periodogram fitting all the $BgcVroi$  photometry simultaneously using the package \texttt{gatspy} \citep{2015ApJ...812...18V,2016ascl.soft10007V}, which generates a Lomb-Scargle \citep{1982ApJ...263..835S} periodogram for a multiband time series (Figure \ref{fig:GLS}). The best period for the combined multiband observations is $\approx$ 12.5 days.

The consistency of the observed periodicity across the time-series was verified through Empirical Mode Decomposition \citep[EMD;][]{1998RSPSA.454..903H,1999AnRFM..31..417H}, which is ideally suited to oscillatory detections in the presence of non-linear and non-stationary processes that may impact the light curves of supernovae. Following the methodology outlined by \citet{2023LRSP...20....1J}, the Intrinsic Mode Functions (IMFs) are extracted from the de-trended time series. Subsequently, a Hilbert-Huang transformation \citep{2008RvGeo..46.2006H} was performed to investigate the instantaneous frequencies across the observing window. A low-order IMF exhibits a frequency associated with a $\approx$12.5~day period for the majority of the time series with little variation. Hence, the EMD processes applied here directly and independently support the GLS periodograms depicted in Figure~{\ref{fig:GLS}}. 

We re-calculate the bolometric light curve using only the \textit{gcroiz}-bands to avoid washing out the periodicity with interpolation. Following identical de-trending methods to the bolometric light curve, we measure the size of the oscillations (peak to trough) to be $\sim$2$\times10^{40}$~erg{\,}s$^{-1}$ over the underlying radioactively-powered flux, which is on the order of 1\% of the peak bolometric luminosity. We perform numerical integration under a single bump in the bolometric light curve to estimate the radiated energy $E_{\rm rad, bump}$ $\simeq 10^{46}$ ergs.

\subsection{Spectra}\label{sec:spec}

The spectroscopic evolution of SN 2022jli is shown in Figure \ref{fig:spectra}, spanning from $-$53 days before maximum light to +48 days after. This includes a spectrum during the first maximum, which is unusual for SNe with a short-lived early peak. The line identifications in this section are based on those by \citet{2009A&A...508..371H}.

The first spectrum obtained during the early excess displays P-Cygni absorption features of Na I D  5891, 5897, and strong Fe II 4924, 5018, 5169 absorption, typical of Type Ic SNe. The broader spectral coverage of the EFOSC2 spectrum (-39 days) reveals Ca II H\&K lines and the Ca II near-IR triplet.

The spectra from +21.5 days show the emergence of a Sc II feature and have a complex blend of narrow emission lines and P-Cygni features like some interacting SNe. The forbidden [Ca II] 7291, 7324 lines are prominent in the later spectra.
We measure the velocities using Gaussian fits to the Fe II 4924, 5018 and 5169 P-Cygni absorption troughs as a proxy for the photospheric velocity. At -39 days we measure $8500\pm300{\rm km}\,{\rm s}^{-1}$, $7000\pm300 {\rm km}\,{\rm s}^{-1}$ at +21 days, reducing further to $6700\pm300{\rm km}\,{\rm s}^{-1}$ by the +48 day spectrum, representing a slow recession of the photosphere inside the ejecta. 

We note the resemblance of the SN 2022jli spectra to SN 1994I but also SN 2004gk although the spectroscopic evolution is not analogous to either object. Given the broad complex at H-alpha, which may be contaminated by Si II and C II, we cannot rule out H in the ejecta or CSM shell(s) but from the line ratios of other H lines expect the contribution of H  to be small. We also fail to identify unambiguous signatures of He in the spectra. 

We include comparison spectra of two representative type Ic events, SN2004gk \citep{2019MNRAS.482.1545S} and SN1994I \citep{2014AJ....147...99M}. The post peak ($+ 25$ days) spectrum of SN2004gk best resembles the $-39$ days spectrum of SN 2022jli but suggest an unusual evolution for to resemble a pre-peak normal Ic spectrum.

We produced a TARDIS \citep{2014MNRAS.440..387K,kerzendorf_wolfgang_2023_8244935} model (Figure \ref{fig:spectra}) with the aim of reproducing the main spectral features of SN 2022jli during first EFOSC2 spectrum. The model is a simple unifrom abundance model with an H- and He-deficient composition dominated by C, O, Si and, Mg and a photospheric velocity of 7,500 \,km\,s$^{-1}$. We adopt a $t_{explosion}$ parameter (time since the start of homologous expansion) of 42 days before the observation at -39 days to best match the observed features. We successfully reproduce the prominent Fe II and Ca II features and continuum shape and show a plausible SN Ic composition can reproduce the spectrum. The model does not reproduce the bump at 6,500 $\rm \AA$ or the emission at 6150 $\rm \AA$. The TARDIS configuration file will be made available as the data behind the figure.
\section{Discussion} 
\label{sec:discuss} 

The data presented in this paper show that SN 2022jli is unusual in many respects. The duration of the initial excess ($\gtrsim 25$ days with no constraining non-detection) is unprecedented for a Type Ic SN. The bolometric light curve peaks at least 59 days after explosion and could be longer given the uncertainty in explosion epoch. In combination with the periodic undulations, the SN 2022jli observational data set is unique.

\subsection{Scenarios for Periodic Variability}
\subsubsection{Interaction with CSM}
\label{sec:csm_shells} 

The bumps we observe in the light curve could be due to ejecta interacting with concentric shells of circumstellar material. During the undulation SN 2022jli is over-luminous by $\sim$ 1 $\times$ 10$^{40}$ \ergs\ for 12.5-days. 
Using the scaling relation $L = \frac{1}{2} M_{\rm CSM} v^{2} / t_{\rm rise}$ \citep{2007ApJ...671L..17S,2007ApJ...668L..99Q,2016ApJ...826...39N}, we estimate the mass required for each bump $M_{\rm CSM,\, bump} \approx 10^{-5} M_{\odot}$, assuming $v=7000\, {\rm km}\,{\rm s}^{-1}$\, from Fe II line velocity measurements, and $t_{\rm rise}$ = 6.3 days.

The average pre-explosion mass-loss rate needed to produce this CSM mass per undulation can be calculated by setting $\dot{M}/{v_{\rm w}}=M_{\rm CSM,bump}/\Delta R$, where $v_{\rm w}$ is the wind velocity and $\Delta R = v t_{\rm bump}$ is the radial distance bounding this CSM mass. For a SN velocity $v=7000$\,km\,s$^{-1}$ and $t_{\rm bump}=12.5$\,days, this gives $\dot{M} \approx{\rm few} \times 10^{-5} (v_{\rm w}/1000\,{\rm km}\,{\rm s}^{-1})$\,M$_\odot$\,yr$^{-1}$. This is consistent with the mass-loss rates observed from Wolf-Rayet stars \citep[e.g.][]{2020MNRAS.499..873S}, suggesting a `typical' wind mass-loss rate could potentially provide the CSM structure needed to explain the periodic undulations of SN 2022jli, if subjected to a periodic modulation. 

Nested shells of dust caused by colliding winds in a 
massive binary system have recently been spectacularly revealed in JWST imaging \cite{2022NatAs...6.1308L}. They showed that the 17 observed shells were due to repeated dust-formation episodes every 7.93 years modulated by periastron passage of 
the companion O5.5fc star in the mutual orbit around the WC7 Wolf-Rayet star.  
An ejection velocity of $v=7000$\kms means the SN shock front travels $\Delta r\simeq54$\,AU in 12.5 days. By comparison, the nested dust shells around WR140 are $\Delta r = 4380\pm120$\,AU. 
If SN 2022jli undulations were due to peaks in CSM density similar to WR140 shells, a 
binary progenitor would need to eject these shells on timescales $\sim$100 times more frequent, 
or with a $\sim$0.2\,yr periodicity. We note that the dust emission in the shells in WR 140 does  not necessairly imply enhanced gas densities \citep{pollock21}.

In this scenario of concentric shells or rings, one might expect that light travel time effects could broaden the undulation timescale as the shock expands. As the SN ejecta hits the back and front of the shells at the same time, the light travel time from the back to front increases as $\Delta t_{\rm lt}\simeq 2v_{\rm ej}t_{\rm exp}/c$, or about 9 days (after 200 days of expansion). While there will be an integrated signal from all parts of the shell, the effect should be to broaden the timescale of the undulations, should the ejecta be photospheric the broadening effect will be less. The data do not clearly support such broadening. 

We can explore potential progenitor candidates using the {\sc BPASSv2.2.2} predictions \citep{bpass2017,bpass2018}, restricting our search to type Ic progenitors (see \citealt{stevance2021}): we select hydrogen deficient systems with surface mass hydrogen $<$0.0001 \msol and surface hydrogen mass fraction $<$0.01. We also restrict our search to helium depleted supernova progenitors, with helium mass fraction $<$0.3. 
We then look for systems that satisfy the period estimate of a WR140-like scenario by searching for P=0.07$\pm0.015$ years (the error is chosen to give a roughly +/-5 days window). 
Finally, we also impose a luminosity and temperature constraint ($\log(L/L_\odot)>5$ and $\log(T/T_\odot>4.3$) as we are looking for WR+O star systems. 
We find 27 BPASS models at solar metallicity that fulfill these requirements, and including the initial mass function weighting we would expect, about 15 such systems to be formed per 1 million M$_\odot$. 
All these systems have primary star (supernova progenitor) masses in a rather narrow range of 10 to 13\,M$_\odot$ while the secondaries range from 24.5 to 60 M$_\odot$. 
Although all these predicted systems have stellar winds around 0.8 $\times 10^{-5}$\,\msol\,yr$^{-1}$, similar to the estimated requirement to create the CSM as mentioned in section \ref{sec:csm_shells}, a key factor in the periodic modulation is the eccentricity of the system. 
Stellar evolution models such as BPASS assume circularised orbits, so we cannot assess how many systems would be born with and maintain sufficient eccentricity.

\subsubsection{Accreting Compact Object}

An alternative mechanism to produce light curve bumps in type Ibc SNe was suggested by \citet{2022MNRAS.517.4544H}. After a stripped 
envelope SN in binary system, the newly born 
neutron star (NS) may receive  a kick in the favourable direction
of its companion. As a result, the NS may penetrate or 
skim the surface of the binary companion. They 
predict that material captured from the companion settles around
the NS and the accretion rate is likely to be super-Eddington. 
The accretion could result in outflows or jets that add further
energy to the SN ejecta and result in additional luminosity. 
Accretion resulting in jets has been modelled by 
\cite{2022MNRAS.516.1846H}, which they propose could power 
bumps in the late time light curves of SNe. Undulations in the 
light curves of SLSNe have been detected 
\citep{2016ApJ...826...39N,2017MNRAS.468.4642I,2021ApJ...913..143G,2022ApJ...933...14H,2023A&A...670A...7W} but no repeating signature has been confirmed over multiple cycles.

\citet{2022MNRAS.517.4544H} calculate that even if only 
$\sim$0.01\,\msol\ is captured by the NS and if only $\sim1$ per cent
of that is accreted, then the energy available would be of order 
$E_{\rm acc}\sim M_{\rm acc}c^2\sim10^{50}$\,erg, which is 
comfortably enough to power a few percent of the total integrated 
SN flux of $E_{\rm rad}\simeq 2.5\times10^{49}$ erg. 
The direct interaction invoked in \cite{2022MNRAS.517.4544H} requires a fairly fine-tuned kick direction and velocity. However a milder interaction as discussed in \cite{2018ApJ...864..119H} and \cite{2021MNRAS.505.2485O} 
may be sufficient. The companion star is inflated by heating from the SN ejecta-companion interaction, and the inflated part of the envelope may interact with the NS, causing periodic accretion on the timescale
of the orbit. 

Accretion powered jets after core-collapse also have 
sufficient energy \citep{2022ApJ...935..108S,2022MNRAS.516.1846H} 
to power the excess flux observed in the light curve of SN 2022jli, 
but a modulation process is required.  In the \citet{2022MNRAS.517.4544H} scenario, the gradual in-spiral of the neutron star into the companion is a pathway for the formation of a Thorne–Żytkow object or TŻO. TŻOs, which are NSs inside an envelope of non-degenerate diffuse material, have been predicted in the literature \citep{1975ApJ...199L..19T,1977ApJ...212..832T} but very few  real candidates exist \citep{2020ApJ...901..135O}. An issue of the \citet{2022MNRAS.517.4544H} scenario is that the orbit of the NS will decay rapidly (within  $\sim$5 orbits), making 15 orbital cycles problematic. However, in the inflated companion case, the low density envelope results in slower orbital decay \citep{2018ApJ...864..119H,2021MNRAS.505.2485O}. 

The accreting compact object model can be thought of as an internal powering source. The energy released must diffuse out through the ejecta on timescales determined by the opacity, density and radius of the   optically thick material. 
\citet{2022ApJ...933...14H} propose that a central origin is disfavoured if the dimensionless depth of the powering source,   $\delta \simeq \frac{t_{\rm bump} \Delta t_{\rm bump} }{t_{\rm rise}^2}$, is significantly less than unity.  With $\Delta t_{\rm bump}\simeq12.5$ and $t_{\rm rise}\simeq60$, then this parameter ranges between 0.2 and 0.7 for the earliest and latest bumps. This would marginally disfavour a central, internal powering source, although \citet{2022ApJ...933...14H} note the expression is quite approximate and should only be treated as an order of magnitude result.

\subsection{Scenarios for Initial Maximum}
\label{subsec:initalmax}

\subsubsection{Interaction with CSM}

MOSFiT modelling of the early excess with an interaction powered model requires $M_{\rm ej}\ \approx 12\,$\,M$_\odot$,  which is compatible with the mass required for the long rise to maximum light. However, the duration requires significant $M_{\rm CSM} (> 3\,$M$_\odot$ in or modelling) which is very large for an SN Ic and would require an exotic mechanism to drive extreme mass-loss shortly before explosion, such as  Pulsational Pair-Instability SN (PPISN) ejections \citep{2017ApJ...836..244W}.

Perhaps the apparent duration of SCE is extended due to enhanced opacity caused by Thomson scattering in the CSM \citep{2012ApJ...756L..22M} which is eventually overtaken by the forward shock. Finally, the red spectrum during the first peak \citep{2022TNSCR1261....1G} would appear inconsistent with luminous circumstellar interaction at early times (i.e., with compact CSM). 

We cannot exclude CSM interaction as the source of the early excess; to do so would require more data to constrain the excess or more sophisticated modelling of the interaction to determine the viability of this scenario. 

\subsubsection{Companion Collision}

Here we consider the emission from the collision of ejecta with the binary companion of SN 2022jli using the model suggested by \citet{2010ApJ...708.1025K}. In this scenario the interaction shocks the SN ejecta, dissipating kinetic energy causing bright optical / UV emission. This additional contribution to the observed luminosity exceeds the radioactively powered SN for a short period, resulting in an early excess.  
We investigate the viability of this model using equations 22 and 23 from \citep{2010ApJ...708.1025K} to estimate the luminosity and the collision luminosity timescale ($t_c$) where ($L_{c,iso} > L_{Ni}$): 

\begin{equation}
    L_{\rm c,iso} = 10^{43}a_{13}M_{\rm c}^{1/4}v_{9}^{7/4}\kappa_{e}^{-3/4}t_{\rm day}^{-1/2} \text{ergs s}^{-1}\ ,
\end{equation}

\begin{equation}
    t_c < 7.3 a_{13}^{2/5} M_{c}^{1/2} v_{9}^{3/10} \kappa_{e}^{1/10} \frac{\kappa_{Ni}}{\kappa_{e}}^{2/5} M_{Ni,0.6}^{-2/5}\ ,
\end{equation}

\noindent where $a_{13} = a / 10^{13}$ cm ($a$ is the orbital separation), $M_{\rm c} =  M / M_{\rm ch}$ is the ejecta mass $M$ in units of the Chandrasekhar mass  ($M_{\rm ch}$), $v_{9} = v_{\rm t} / 10^9$ cm s$^{-1}$, $v_{\rm t} = 6\times10^{8} \zeta_v (E_{51} / M_c) ^{1/2} $ cm s$^{-1}$ (following \citealt{2010ApJ...708.1025K}, we adopt $\zeta_v = 1.69$), $\kappa_{Ni}$ is the opacity in the \Ni\ dominated region and $\kappa_{e}$ is the ejecta opacity outside this region. The time since explosion is given as\ $t_{\rm day}$,  $M_{\rm Ni,0.6} = M_{\rm Ni} / 0.6 M_\odot$, $E_{51} = E / 10^{51}$ \ergs, $E$ is the explosion energy. We adopt an indicative ejecta mass of 12 $M_\odot$ from MOSFiT modelling in section \ref{sec:lightcurve_model}  and set $\kappa_{\rm e}$ = $\kappa_{\rm Ni} = 0.1$ cm$^2$ g$^{-1}$. We set $M_{\rm Ni} \approx 0.7$, (for $f_{\rm Ni} \approx 0.06$ from MOSFiT) and set $v_{\rm t}$ = 8,500 $\rm km s^{-1}$ from direct measurement of the -29 day spectrum (during the early excess). 

To produce the observed early luminosity on the order of $\sim 10^{42}$ ergs s$^{-1}$ and timescale of $\sim$ 10 days, we would need to have separation $\sim$ 1 AU. For these parameters we calculate $t_{\rm c} \lesssim 17.5$ days and $L_{\rm c,iso} \approx 8 \times 10^{43}$ \ergs\ for $t_{\rm d}$ = 2 days and $L_{\rm c,iso} \approx 2 \times 10^{42}$ \ergs\ for $t_{\rm d}$ = 20.

With these assumptions for separation and ejecta mass our observations are compatible the direct collision of the ejecta with the companion star. It is important to note this scenario requires a favourable viewing angle, \citet{2010ApJ...708.1025K} predict that the collision should be visible in only $ \sim 10 \%$ of cases and that an orbital separation of $\sim$ 1 AU at the time of interaction may require the object to be close to pericenter. An important additional qualification of the \citet{2010ApJ...708.1025K} model calculation is the assumption that the companion to the exploding star is filling its Roche lobe, as in a thermonuclear binary star explosion. Therefore, this calculation should be regarded only as an illustrative estimate of the energetics of a companion interaction. The progenitor systems considered in \ref{sec:csm_shells} have separations of 0.5 AU and upwards which is  compatible with the 1 AU separation adopted for this calculation, although we note that they are typically not filling their Roche lobe and interacting. Should the excess be powered by companion collision one might expect to observe late-time H$\alpha$ emission from the companion and at late times observe a surviving but inflated companion star.

\section{Conclusions}

We have presented detailed, multi-wavelength, high cadence observations of the unprecedented Type Ic SN 2022jli. We attribute the long rise to maximum as the signature of a large ejecta mass ($M_{\rm
ej}\approx12\pm6$M$_{\odot}$). Future nebular phase spectroscopy may provide an independent estimate of the core mass from the [Ca II] and [O I] line ratios \citep{1989ApJ...343..323F}. 

We provide the first unambiguous detection of periodic behaviour in a SN optical light curve, measuring a period of $\sim$12.5 days and amplitude $\sim$1\% of the SN maximum light, repeating over a time window of at least $\sim$ 200 days. This could be explained by discrete episodes of shock heating from interaction with a structured CSM produced through modulated mass-loss of the progenitor star in a binary system. We also consider companion-compact object interaction as the energy source but favour a structured CSM.

We also observe a prolonged early excess and consider two scenarios: CSM interaction and ejecta-companion interaction. Based on the methods presented in this work we cannot distinguish between these two scenarios. A dense CSM shell requires several M$_\odot$ of material around the progenitor star requiring exotic phenomena like PPI \citep{2017ApJ...836..244W} shortly before explosion. Although only visible in 10\%\ of cases we cannot rule out ejecta-companion interaction, especially given that binarity is already invoked to explain the periodic undulations. However, this scenario has strict requirements on explosion energy and binary separation. 

 SN 2022jli is the subject of further study and multi-wavelength observations (Moore et al., in preparation). Late-time high resolution JWST or HST photometry may reveal the origin of the CSM or a surviving, inflated companion star \citep{2015A&A...584A..11L,2018ApJ...864..119H}.

%\begin{acknowledgments}
\vspace{1cm}
ATLAS is primarily funded through NASA grants NN12AR55G, 80NSSC18K0284, and 80NSSC18K1575. The ATLAS science products are provided by the University of Hawaii, Queen’s University Belfast, STScI, SAAO and Millennium Institute of Astrophysics in Chile. MN, SS, AA, and XS are supported by the European Research Council (ERC) under the European Union's Horizon 2020 research and innovation programme (grant agreement No.~948381) and by UK Space Agency Grant No.~ST/Y000692/1. Lasair is supported by the UKRI Science and Technology Facilities Council and is a collaboration between the University of Edinburgh (grant ST/N002512/1) and QUB (grant ST/N002520/1) within the LSST:UK Science Consortium. ZTF is supported by National Science Foundation grant AST-1440341 and a collaboration including Caltech, IPAC, the Weizmann Institute for Science, the Oskar Klein Center at Stockholm University, the University of Maryland, the University of Washington, Deutsches Elektronen-Synchrotron and Humboldt University, Los Alamos National Laboratories, the TANGO Consortium of Taiwan, the University of Wisconsin at Milwaukee, and Lawrence Berkeley National Laboratories.  This work is based on observations collected at: the European Organisation for Astronomical Research in the Southern Hemisphere, Chile, as part of ePESSTO+ (the advanced Public ESO Spectroscopic Survey for Transient Objects Survey). ePESSTO+ observ'ations were obtained under ESO program ID 108.220C (PI: Inserra).
The Las Cumbres Observatory,   LCO data have been obtained via an OPTCON proposal (IDs: OPTICON 22A/004, 22B/002; European Union’s Horizon 2020 grant agreement No 730890) and the LCO team is supported by NSF grants AST-1911225 and AST-1911151.  SS, SAS and SJS acknowledge funding from STFC Grants ST/X006506/1 and ST/T000198/1. DBJ and SDTG acknowledge funding from STFC grant awards ST/T00021X/1 and ST/X000923/1. DBJ and WB acknowledge support from the Leverhulme Trust via the Research Project Grant RPG-2019-371. LG and CPG acknowledge financial support from the Spanish Ministerio de Ciencia e Innovaci\'on (MCIN), the Agencia Estatal de Investigaci\'on (AEI) 10.13039/501100011033, and the European Social Fund (ESF) ``Investing in your future" under the 2019 Ram\'on y Cajal program RYC2019-027683-I, the Marie Sk\l{}odowska-Curie and the Beatriu de Pin\'os 2021 BP 00168 programme and the PID2020-115253GA-I00 HOSTFLOWS project, from Centro Superior de Investigaciones Cient\'ificas (CSIC) under the PIE project 20215AT016, and the program Unidad de Excelencia Mar\'ia de Maeztu CEX2020-001058-M. We acknowledge funding from ANID, Millennium Science Initiative, ICN12\_009. GL is supported by a research grant (19054) from VILLUM FONDEN. TWC thanks the Yushan Young Fellow Program by the Ministry of Education, Taiwan for the financial support.

%\end{acknowledgments}

\facilities{ATLAS, NTT, ZTF, ASAS-SN, LCO, GAIA, SWIFT}

\software{
Astropy \citep{Astropy2013,Astropy2018,astropy2022},
Numpy \citep{Harris2020}, 
Matlpotlib \citep{2018ApJS..236....6G},
Mosfit \citep{2018ApJS..236....6G},
Hoki \citep{hoki},
PSF \citep{2023arXiv230702556N}
}

\bibliography{bib}
\bibliographystyle{aasjournal}

\appendix 
\section{MOSFiT Posterior}

\begin{figure}[h!]
    \centering
    \includegraphics[width=0.85\linewidth]{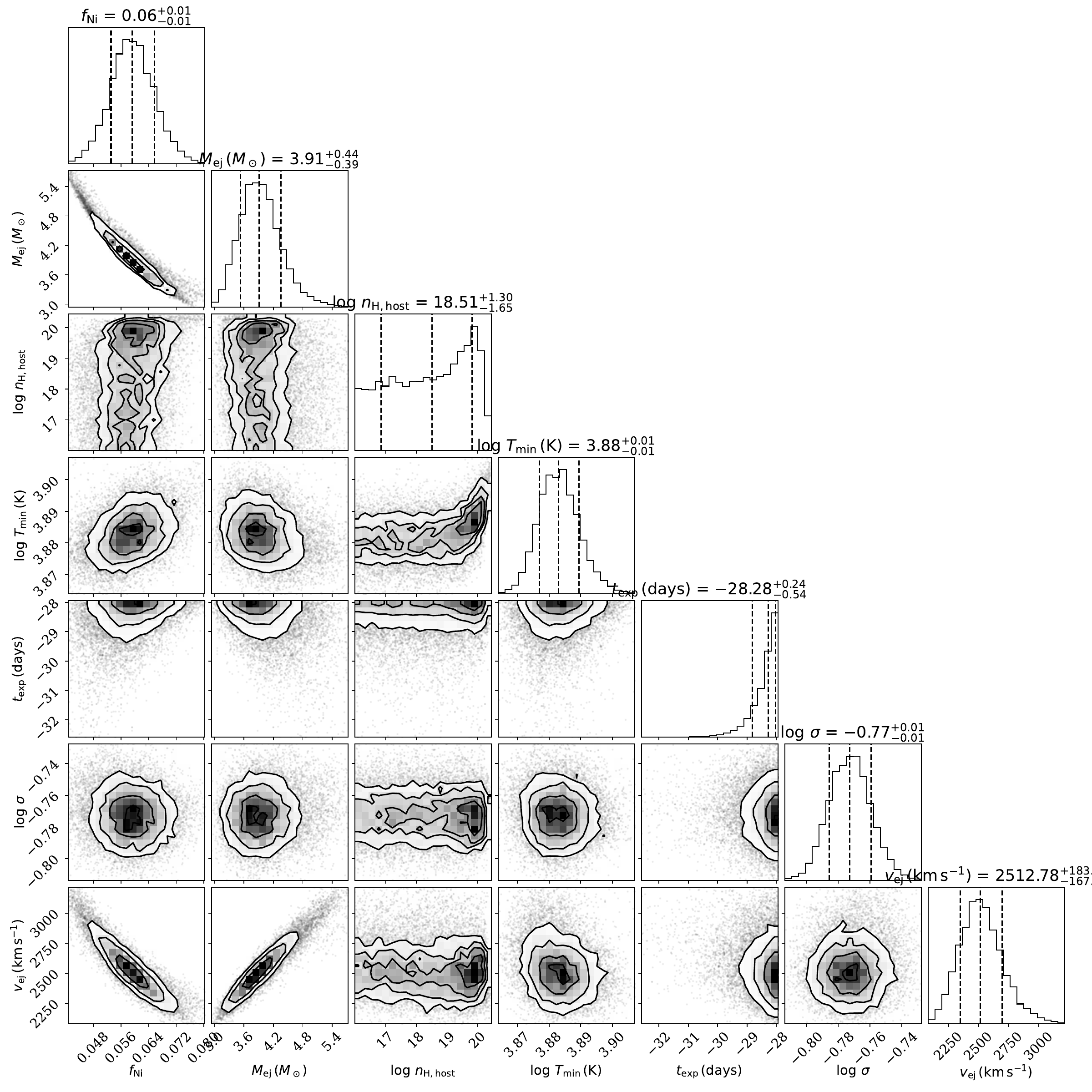}
    \caption{Posterior distribution for the physical parameters of the MOSFiT default model fit to SN 2022jli. In this model $f_{\rm Ni}$ is the fraction of \Ni\ in the ejecta and M$_{\rm ej}$ is the mass in solar masses. Ejecta velocity is $V_{\rm ej}$ and $log\ n_{\rm, host}$ is the logarithm of the host H column density. $T_{\rm min}$ is the temperature floor as defined in \citet{2017ApJ...850...55N}, $t_{\rm exp}$ determines the explosion epoch relative to the earliest photometry point. The $sigma$ parameter ($\sigma$) is a white noise parameter which when added to all data gives a $\chi^2$ equal to 1.} 
    \label{fig:corner}
\end{figure}

We the show the posterior distribution in Figure \ref{fig:corner} for the \Ni\ model fit to SN 2022jli described in section \ref{sec:lightcurve_model}

\end{document}